# Position and Mode Dependent Optical Detection Back-Action in Cantilever Beam Resonators

T. Larsen[1,2], S. Schmid[1,3], S. Dohn[1], J. E. Sader[4], A. Boisen[1] and L. G. Villanueva[1,2]

[1]*DTU Nanotech, Technical University of Denmark (DTU), Lyngby, DK-2800, Denmark*
[2]*Ecole Polytechnique Fédérale de Lausanne (EPFL), Lausanne CH-1015, Switzerland*
[3]*Institute of Sensor and Actuator Systems, Vienna University of Technology, 1040 Vienna, Austria*
[4]*School of Mathematics and Statistics, The University of Melbourne, Victoria 3010, Australia*

**ABSTRACT**

Optical detection back-action in cantilever resonant or static detection presents a challenge when striving for state-of-the-art performance. The origin and possible routes for minimizing optical back-action have received little attention in literature. Here, we investigate the position and mode dependent optical back-action on cantilever beam resonators. A high power heating laser (100 µW) is scanned across a silicon nitride cantilever while its effect on the first three resonance modes is detected via a low-power readout laser (1 µW) positioned at the cantilever tip. We find that the measured effect of back-action is not only dependent on position but also the shape of the resonance mode. Relevant silicon nitride material parameters are extracted by fitting the temperature-dependent frequency response of the first three modes to finite element (FE) simulations. In a second round of simulations, using the extracted parameters, we successfully fit the FEM results with the measured mode and position dependent back-action. Finally, different routes for minimizing the effect of this optical detection back-action are described, allowing further improvements of cantilever-based sensing in general.

## I. INTRODUCTION

Micro- and nanoscale mechanical structures have, over the past decades, shown to be an extremely versatile sensing platform for a large number of applications [1-3]. Particular examples include the inertial imaging of particles [4], real-time monitoring of cellular mass gain [5], detection of molecular interaction forces [6-8], material properties characterization [9], and probably the most important of all: atomic force microscope (AFM) [10-12]. The sensing principle is typically based on monitoring the displacement or the resonance frequency of the mechanical device, either of which is altered due to the physical or chemical magnitude that is under study. The latter method usually provides a better resolution and general performance. However, in either case, the motion of the mechanical device needs to be measured. Thus, the foundation of these systems consists of a mechanical device and a transduction method for its motion (both actuation and detection).

Various transduction methods have been used over the years, e.g. piezoresistive [13], piezoelectric [14], magnetomotive [15], electrostatic [16], and optical [17]. Typically, the goal is to perform this transduction with as high resolution as possible to be able to detect the displacements with as large as possible signal-to-noise ratio. This normally comes at a price, as the back-action of the measurement technique into the mechanical resonator increases with the displacement resolution. In fact this can be traced down to quantum physics [18], but we can also find a more classical explanation. To improve sensitivity, it is typically necessary to increase the probing power, which means that relative fluctuations in said power are larger in absolute value, which therefore causes a larger effect on the resonator. In some cases, back-action effects are preferred either to cool the resonator or to observe quantum forces [19-22]. In many cases, however, back-action feeds noise into the system [23-26].

If we focus on the particular case of mechanical resonators measured via optical detection, as for example AFM or cantilever arrays for biosensing, there are two major ways in which laser fluctuations affect the resonance frequency of the mechanical device: radiation pressure and bolometric effect. On the one hand, the radiation pressure exerted by the laser beam can depend on the displacement of the resonator and thus affect its resonance frequency. On the other hand, more intuitively, laser light absorption increases the temperature of the resonator (bolometric effect). This increase in temperature changes material properties and dimensions [27], which are relatively easy to calculate in order to estimate the associated change in the resonance frequency. Finally, as a second order effect of the change in dimensions, there will be some stress generated in the resonator and a curvature might be induced both in the longitudinal and transversal directions. These two latter points have been receiving some attention in the last few years [27-31]. For clamped-clamped structures the effect of stress is clearly understood, but for clamped-free structures (i.e. cantilevers) it has been a pivotal point of discussion since 1975 [32, 33].

The initial paper from Lagowski was quickly followed by a series of contradicting theoretical papers. More elaborate theories have more recently been developed by Sader et al. [28, 29] and by Tamayo et al. [27, 34]. First, Sader developed a theoretical model for the effect of (surface) stress on the stiffness of monomaterial cantilevers, even showing the very interesting possibility of finding buckling for *tensile* stress [35]. The conclusion from Sader's original model is that the stiffness can be tuned in cantilevers via the loss of symmetry close to the clamp, where the applied stress cannot be totally released. It predicts that the relative change in frequency scales with $w^3/Lt^2$, where $w, L, t$ are width, length and thickness respectively; i.e. the thinner the structure, the larger the effect. This is one of the reasons that motivated Tamayo's initial work using bimorph cantilevers [30]. In this case, the temperature induces not only a stress distribution close to the clamp but also a curvature, which typically dominates in the case of cantilevers designed to be used in surface stress measurements (thin, long, and with asymmetric mechanical configuration) [36]. Relative changes in torsional mode frequencies of up to 40% have been observed for a readout laser power of 1.7 mW. More recently, Nieva et al. [37] have also shown the importance of curvature on the resonance frequency of bimorph cantilevers. However, the majority of cantilever beams are monomaterial, as it is the case for Si cantilevers for AFM.

In this work we study the effect of the laser position on the resonance frequency of a monomaterial cantilever. A heating laser (100 µW) is used to locally heat the cantilever, while a low power readout laser (1 µW) is positioned at the cantilever tip. We show that positive and negative changes in the resonance frequency can't be explained solely by temperature induced changes in mechanical properties but the effect of a local stress gradient has to be accounted for, too.

**II. EXPERIMENTALS**

We select one of the most common materials in MEMS for the production of our structures: non-stoichiometric (silicon-rich) silicon nitride (SiNx). Cantilever structures are fabricated using a simple 2 step lithography process, see Fig 1A. Silicon oxide is grown on a silicon wafer via thermal oxidization. Silicon-rich silicon nitride is subsequently deposited using Low Pressure Chemical Vapor Deposition (LPCVD). Using photoresist as etch mask, reactive ion etching is used to pattern the silicon nitride on the front and backside of the wafer. Anisotropic KOH etching from the backside while mechanically protecting the front side followed by HF etching are used to release the cantilevers. An optical micrograph of the released cantilevers can be seen in Fig. 1B. The cantilevers have similar dimensions to other examples found in the literature with thickness ($t$) of 500 nm, width ($w$) of 100 µm and length ($L$) between 460 µm and 630 µm. The characterization of the cantilevers is performed in vacuum ($P \leq 10^{-5}$ mbar), at room temperature. An external piezoelectric shaker is used for the actuation and a Polytec laser-Doppler vibrometer to detect the motion. In order to have a more stable experiment, we use the detecting laser with minimum power (1 µW) focused at the free end of the cantilever (Fig. 1B – bright spot). The resonance frequency is monitored *simultaneously* for the three first out-of-plane flexural modes. Two types of experiment were performed: i) uniform heating of the cantilever and ii) local heating with the heating laser. The former case is performed by using a PID controlled Peltier element to heat the cantilever chip. The steady state frequency is determined, which practically corresponded to a 99% convergence to the final value. In the latter case, the heating laser (100 µW) is moved along the cantilever surface (Fig. 1B, inset, dark spots). At each of those points the laser is alternatively switched on-off at a rate of 0.2 Hz in order to perform a differential measurement of the heating effect.

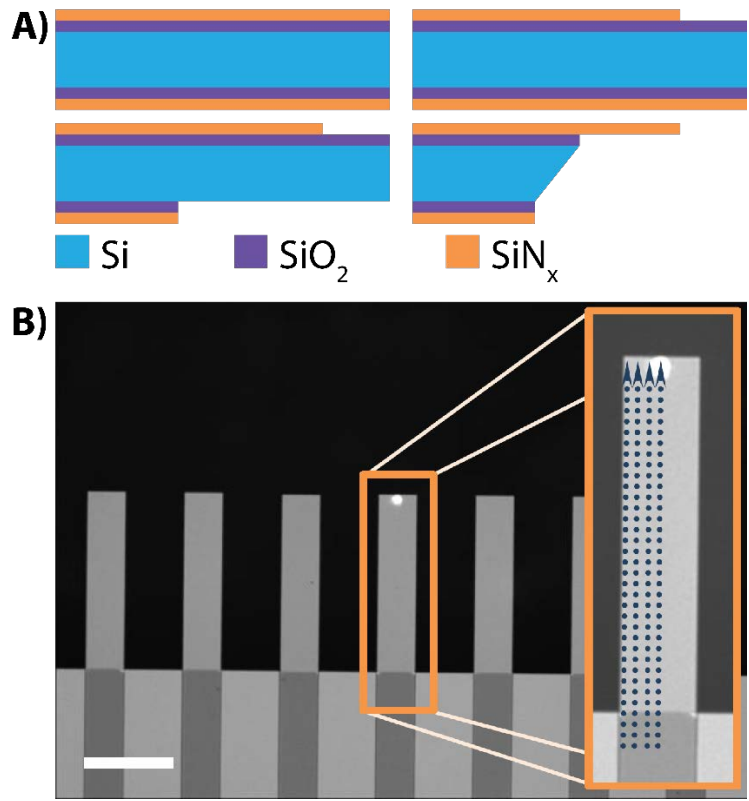

Figure 1: A) Schematic of the cantilever fabrication process. Silicon oxide is grown by thermal oxidation and silicon nitride deposited via LPCVD on a silicon wafer. Patterning of the layers is obtained via reactive ion etching using photoresist as etch mask. Anisotropic KOH etch, with mechanical front side protection, is used to release the cantilevers from the back. Membranes of oxide is finally removed in HF. B) Optical micrograph of the fabricated cantilevers. Scale bar, 250 µm. The Doppler vibrometer detecting laser (1 µW) is focused at the free-end of the cantilever to be tested. A second laser (heating laser, 100 µW) is then scanned through the cantilever (see dotted arrows) while being turned on-off with a frequency of 0.2 Hz in order to provide differential measurements.

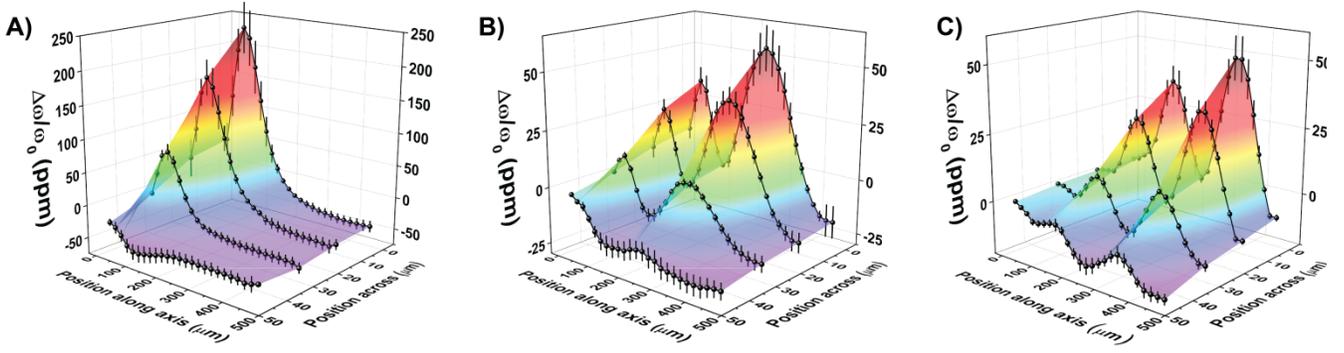

Figure 2: Relative frequency shift (in parts per million, ppm) of the three first out-of-plane flexural modes of the cantilevers shown in Fig. 1. Scattered data corresponds to the average of the measurements over 4 different cantilevers and the error bars correspond to twice the standard deviation of the measurements. Surface color maps and connecting lines are intended as guidance for the eye.

**III. RESULTS AND DISCUSSION**

We first analyze the frequency shift due to a uniform heating to gain some insight into the material properties of the ensemble we use. The experimental results (not shown) provide a relative frequency change over temperature for the first, second and third flexural mode of $-350$ ppm/K, $-90$ ppm/K, and $-60$ ppm/K, respectively, where ppm stands for parts per million. As mentioned above, the uniform heating changes material properties, dimensions and, as a second order effect, it introduces stress near the clamping region of the cantilever. It is evident that this latter effect (stress tuning) has to be taken into account, as changes in material properties or dimensions would yield mode-independent relative changes. By fitting FEM results to our measurements we obtain the following set of material properties for $SiN_x$: $\frac{\Delta E}{E}/\Delta T \approx -50$ ppm/K (temperature dependence of Young's modulus), CTE $\approx 2.5$ ppm/K (coefficient of thermal expansion), and $\sigma_0 \approx 200$ MPa (tensile intrinsic stress).

Then we analyze the effect of local heating by a laser. Thermal expansion caused by the heating laser generates localized stress peaking at the laser spot position. The observed relative frequency shifts for the first three flexural modes as a function of heating laser position are plotted in Fig. 2 for the case of a 460 μm long cantilever. A similar trend is observed for a 630 μm long cantilever. The shifts for each mode are relative to the frequency measured with no heating. The results shown in Fig. 2 evidence a non-uniform response across the cantilever that is fundamentally different depending on the mode shape. The higher the resonant mode, the more complex the response becomes when moving the heating source along the cantilever. The observed dependence on the resonance mode, e.g. one mode shows a negative change while another a positive, can again not solely be explained by temperature induce changes in material properties.

Using the material properties extracted from the previous experiment, we perform another round of FE simulations to reproduce the laser-heating experiment (see Fig. 3). The resonance frequencies of the first three modes are simulated while moving a heat source along and across the cantilever surface (figure only shows the results for the central line). The simulations capture the trend for all three modes, showing both position and mode dependences, and yielding a thermal conductivity for the utilized silicon-rich silicon nitride of $\kappa \approx 150 \frac{W}{m \cdot K}$. Again, the simulation results cannot solely be explained via material properties or dimensional changes.

Therefore, the presented results provide further evidence that (surface) stress affects the cantilever resonance frequencies. In addition, the results show that such stress effects are dependent on the spatial distribution and on the resonant mode number. Previous reported experimental studies have, to our knowledge, not revealed details at this level [27, 30, 31]. We think that the main reason why we are able to observe this mode-shape dependence is because our cantilevers are monomaterial, i.e. we do not have any curvature effects. We think that this effect is related to the interaction of thermally generated local strain/stress distribution and the mechanical stress associated to the motion (proportional to the second derivative of the mode shape).

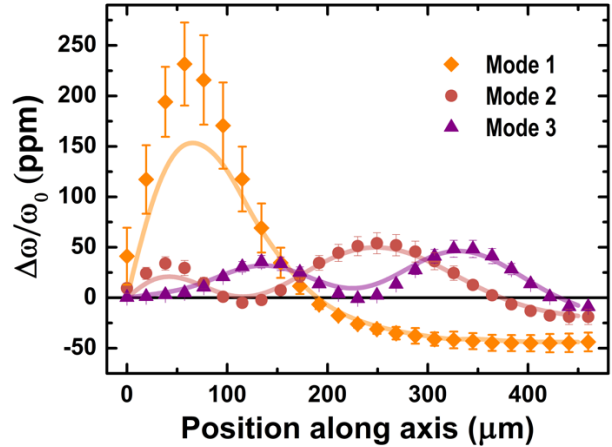

Figure 3 Comparison between experimentally measured (scattered data with error bars) relative frequency shift upon laser illumination, together with FEM simulated results (lines) for each one of the three first out-of-plane flexural modes. Data is taken from Figure 2, for the laser located along the center line of the cantilever.

It is important to emphasize the significance of the results in Figs. 2-3, as they provide a roadmap to minimize optical detection back-action on the resonance frequency of a cantilever during sensing experiments. In order to optimize transduction efficiency, the preferred locations for the laser spot on a cantilever are typically those where either the displacement or its first derivative is maximized. This is done so that the signal-to-noise ratio is maximized and thus the noise in the determination of the frequency is minimized [25]. However, fluctuations of the laser power or position can directly feed into the system as frequency noise. Consequently, the position-dependent sensitivity to optical detection back-action has to be taken into account when choosing the readout laser position for an optimum operation.

For the sake of illustrating our point, let us assume an optical detection setup that has two sources of noise affecting the transduction of the motion, i.e. system noise (e.g. shot noise, amplifier noise, laser power fluctuations, etc.) and thermomechanical noise. Let us also assume that the system noise level is 10 times smaller than the thermomechanical noise when the latter is maximum. Figure 4 shows the arbitrarily scaled Allan Deviation (a magnitude quantifying the frequency noise) for the fundamental mode as a function of the laser spot location when considering only the system noise, the thermomechanical noise or the combination of both of them. Then, on top of this, let us assume that the laser power is not completely stable. In that case, as discussed above, we would see frequency jittering that would be directly caused by the fluctuations of the laser power. Figure 4 also shows the Allan Deviation that results from taking such power variations into account for two (arbitrary) levels of back-action. As can be seen, the laser location for an optimized frequency stability changes when optical back-action can't be neglected. A similar reasoning can be used when the experimental setup is prone to vibrations that affect the laser position along the cantilever. In this latter case, however, back-action effects are proportional to the spatial derivative of the curves shown in Fig. 3, thus yielding a different location of the optimum readout position.

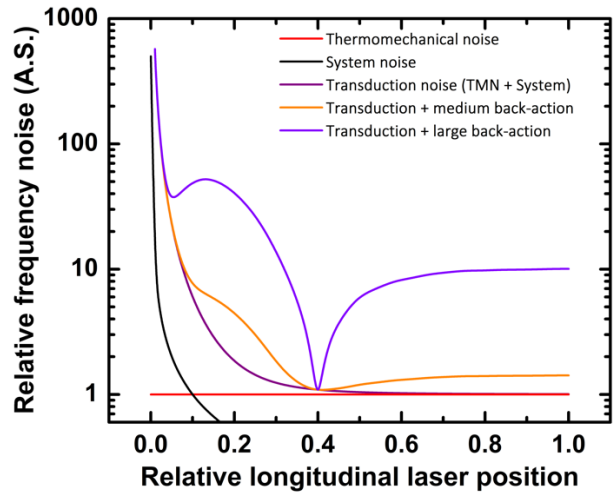

Figure 4 Plot showing arbitrarily scaled Allan Deviation as a function of the laser position along the cantilever for the fundamental mode (scaled to the total length). The result coming from thermomechanical noise and system noise is shown separately and combined into what we call transduction noise. Finally, the contribution from laser power fluctuations is accounted for with two magnitudes for the back-action effect. The laser location at which the frequency stability is optimized depends strongly on the back-action effect.

## IV. CONCLUSION

In conclusion we have observed that the laser beam on a cantilever can lead to positive or negative changes in resonance frequency depending of the positioning of the laser beam and the specific resonant mode of interest. The observed behavior cannot be explained purely by temperature-induced changes in mechanical properties. Global curvature of the structures does not suffice either, as our structures are monolayer and thus curvature effects are suppressed. The effect of local stress gradients has to be taken into account, and this is evident in the observed mode-shape dependence of our results. Using two rounds of FE simulations and matching them to our experimental results we are able to extract the cantilever's material properties. These results provide a roadmap to minimize frequency noise. Importantly, the effects of back-action should be quantified in each experimental setup in order to be able to optimize frequency resolution. Our findings put into perspective most of long-standing assumptions about the optimal location of a laser spot when characterizing small mechanical devices.

## ACKNOWLEDGEMENTS

The authors would like to thank financial support from the Swiss National Science Foundation (PP00P2-144695), the European Commission (PCIG14-GA-2013-631801) and the Australia Research Council grants scheme.